\documentclass[aps,twocolumn,prl,preprintnumbers,amsmath,amssymb]{revtex4-1}
\allowdisplaybreaks[4]
\newcommand{\lsim} 
 {\ \raise.35ex\hbox{$<$}\kern-0.75em\lower.5ex\hbox{$\sim$}\ }
\newcommand{\gsim}
 {\ \raise.35ex\hbox{$>$}\kern-0.75em\lower.5ex\hbox{$\sim$}\ }

\usepackage{graphicx}
\usepackage{dcolumn}
\usepackage{bm}
\usepackage{amsmath}
\usepackage{times}
\usepackage[usenames]{color}
\usepackage{natbib}
\usepackage{ulem}

\begin{document}
\title{Perovskite as a spin current generator}
\author{Makoto Naka$^{1}$, Yukitoshi Motome$^2$ and Hitoshi Seo$^{3,4}$}
\affiliation{$^1$Waseda Institute for Advanced Study, Waseda University, Shinjuku, Tokyo 169-8050, Japan}
\affiliation{$^2$Department of Applied Physics, The University of Tokyo, Bunkyo, Tokyo 113-8656, Japan}
\affiliation{$^3$Condensed Matter Theory Laboratory, RIKEN, Wako, Saitama 351-0198, Japan}
\affiliation{$^4$Center for Emergent Matter Science (CEMS), RIKEN, Wako, Saitama 351-0198, Japan}
\date{\today}
\begin{abstract}
We theoretically show that materials with perovskite-type crystal structures provide a platform for spin current generation, taking advantage of a mechanism requiring neither the spin-orbit coupling nor a ferromagnetic moment, but is based on spin-split band structures in certain kinds of collinear antiferromagnetic states.
By investigating a multiband Hubbard model for transition metal compounds by means of the Hartree-Fock approximation and the Boltzmann transport theory, we find that a pure spin current is induced by an electric field applied to a C-type antiferromagnetic metallic phase.
The spin current generation originates from a cooperative effect of spatially anisotropic electron transfer integrals owing to the GdFeO$_3$-type lattice distortion, which is ubiquitous in many perovskites, and the collinear spin configuration. 
We discuss our finding from the symmetry point of view, in comparison with other spin current generator candidates with collinear antiferomagnetism.
We also propose several ways to detect the phenomenon in candidate perovskite materials. 
\end{abstract} 


\maketitle
\narrowtext



%
%

%


Perovskites, a large family of compounds with chemical formula {\it ABX}$_3$ as the mother phase and their related structures, show versatile functionalities and serve as one of the most well-studied textbook materials in condensed matter physics~\cite{imada, cheong, maekawa, tilley}.
In particular, TM-based perovskites exhibit a wide range of properties, e.g., ferroelectricity, metal-to-insulator (MI) transition, magnetoelectric effect, spin crossover phenomenon, superconductivity, and photovoltaic effect~\cite{wainer, torrance, kimura, korotin, bednorz, kojima}.
The diverse physical properties are created by chemical substitutions which controls the electronic states by tuning their band filling, bandwidth, and dimensionality. 
Amazingly, the rich variety originates from a common framework of $d$ electrons of the TM elements {\it B} and $p$ electrons. 
Here we show that yet another functionality can be extracted, which has been overlooked for decades: spin current generation useful for spintronic devices. 

In the field of spintronics, search for materials for efficiently generating spin currents has been pursued, not limiting to conventionally-used ferrimagnets and ferromagnets~\cite{fert, slonczewski, tserkovnyak, uchida}. 
Semiconductors and metals with heavy element have been considered as candidates by the realization of the spin Hall effect~\cite{dyakonov, hirsch, murakami, sinova, kato, wunderlich, saitoh}, whose mechanism relies on the spin-orbit coupling. 
Recently, the search has been extended to antiferromagnets with noncollinear~\cite{zelezny, kimata} and collinear magnetic structures~\cite{naka, hayami, gonzalez, yuan, hayami2}.
For future applications, exploring such a functionality in a feasible way using ubiquitous materials is crucial. 

\begin{figure}[t]
\begin{center}
\includegraphics[width=1.0\columnwidth, clip]{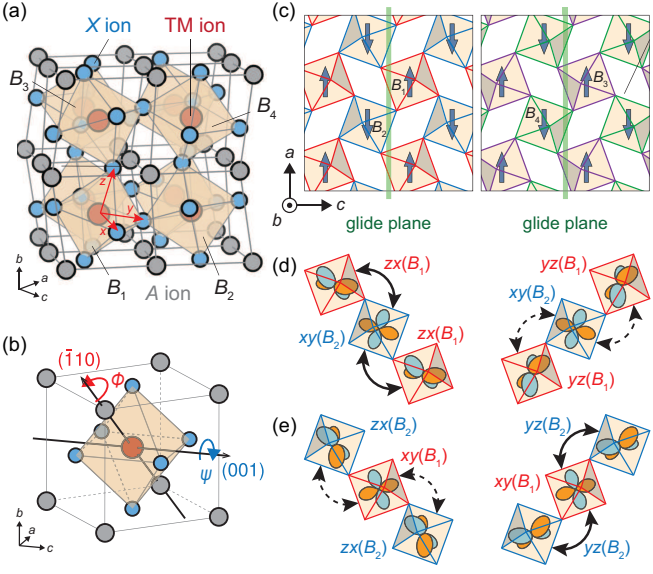}
\end{center}
\caption{
(a) Perovskite structure with the GdFeO$_3$-type distortion. 
$B_1$-$B_4$ denote the {\it BX}$_6$ octahedra contained in the unit cell, connected by symmetry operations thus crystallographically equivalent. 
The $abc$ axes are defined for the {\it Pnma} space group.
The $xyz$ axes represent the local coordinate defined for each octahedron.
(b) Two kinds of {\it BX}$_6$ rotation modes of the GdFeO$_3$-type distortion described in the text. 
(c) Schematic illustration of the C-type AFM spin configuration in the two $ac$ planes.
The blue arrows represent up and down spin moments of the $d$ orbitals in the TM ions.
The green lines denote the glide plane perpendicular to the $ac$ plane.
Schematic illustrations of the anisotropies of the $t_{2g}$-$t_{2g}$ transfer integrals on (d) $B_1$-$B_2$-$B_1$ and (e) $B_2$-$B_1$-$B_2$ bonds along the $(101)$ and $(10\bar{1})$ directions. 
The magnitude of the transfer integrals denoted by the solid arrows are larger than those by the dashed arrows in the presence of the distortion.
}
\label{fig1}
\end{figure}
In this Letter, we theoretically propose that perovskite TM compounds work as an efficient spin current generator. 
Adopting a multiorbital Hubbard model for the $d$ orbitals of the TM ions, we investigate spin transport properties in an applied electric field.
A key ingredient is the so-called GdFeO$_3$-type lattice distortion~\cite{glazer, okeeffe}, commonly seen in the {\it ABX}$_3$ compounds and known to control the bandwidth via the tolerance factor, a ratio between the ionic radii of three elements. 
The distorted perovskite structure is shown in Fig.~\ref{fig1}(a), in which the regularly aligned {\it BX}$_6$ octahedra on the cubic lattice rotate so as to fill the crystal voids around the {\it A} sites, as illustrated in Fig.~\ref{fig1}(b) (see below). 
Consequently the symmetry of the crystal structure is lowered to an orthorhombic {\it Pnma} space group.  
We will show that the spin current conductivity becomes nonzero in a metallic state with a C-type AFM (C-AFM) ordering illustrated in Fig.~\ref{fig1}(c) and increases with the GdFeO$_3$-type distortion. 
The spin current generation here originates from the spin splitting of the energy band in the C-AFM state, due to the spatial anisotropies of the transfer integrals arising from the GdFeO$_3$-type distortion. 
We discuss the symmetry aspect of the spin current generation in comparison with the previous studies for collinear antiferromagnets.
Finally, we propose how to verify our proposal by presenting candidate perovskite TM compounds.

Let us introduce how to incorporate the GdFeO$_3$-type distortion for modeling {\it ABX}$_3$.
We start with the tight-binding $d$-$p$ model composed of the five $d$ orbitals of the TM ions with the cubic crystalline field splitting together with the three $p$ orbitals of the ligands~\cite{mizokawa, mochizuki2, mochizuki}.
The GdFeO$_3$-type distortion is characterized by two kinds of rotation modes of the {\it BX}$_6$ octahedra~\cite{okeeffe}: a major rotation $\pm\phi$ around the $(110)$ or $(\bar{1}10)$ axis followed by an additional tilting $\psi=\pm\arctan [\sqrt{2}(1-\cos \phi)/(2+\cos \phi)]$ around the $(001)$ axis [see Fig.~\ref{fig1}(b)].

Here, we adopt an effective tight-binding model for the $d$ orbitals derived by the second-order perturbation in terms of the $d$-$p$ electron transfer integrals~\cite{maekawa, mochizuki2}, which is given by 
\begin{equation}
{\cal H}_{\rm 0} = 
\sum_{i \beta \sigma} \epsilon_{\beta} n_{i \beta \sigma} + \sum_{ij \beta \beta' \sigma} [\hat{t}^{dd}_{ij}(\phi)]_{\beta \beta'} c^{\dagger}_{i \beta \sigma} c_{j \beta' \sigma},
\end{equation}
where $c_{i \beta \sigma}$ and $n_{i \beta \sigma}(=c^{\dagger}_{i \beta \sigma}c_{i \beta \sigma})$ are the annihilation and the number operators of an electron with spin $\sigma$ of the $d$ orbital $\beta$($=x^2-y^2, 3z^2-r^2, xy, yz, zx$), respectively, represented in the local $xyz$ axes fixed on the $i$th octahedron shown in Fig.~\ref{fig1}(a), and the spin axes are globally defined for all the sites.
We assume that the energy levels of the $d$ orbitals are given as $\epsilon_{x^2-y^2} = \epsilon_{3z^2-r^2} = 3 \Delta_{\rm cf}/5$ and $\epsilon_{xy} = \epsilon_{yz}= \epsilon_{zx} = - 2 \Delta_{\rm cf}/5$ with the octahedral crystalline field splitting $\Delta_{\rm cf}$ between the $e_g$ and $t_{2g}$ manifolds.
We consider the nearest-neighbor $d$-$d$ transfer integral, given by  
\begin{equation}
[\hat{t}^{dd}_{ij}(\phi)]_{\beta\beta'} = - \frac{1}{\Delta_{\rm ct}} \sum_{\gamma \gamma'} [\hat{\tau}^{pd}_{i;ij}]^\top_{\beta \gamma} [\hat{\cal R}_{ij}(\phi)]_{\gamma \gamma'} [\hat{\tau}^{pd}_{j;ij}]_{\gamma' \beta'},
\end{equation}
where $\Delta_{\rm ct}$ is the charge transfer energy between the $p$ and $d$ orbitals. 
$\hat{\tau}^{pd}_{i;ij}$ is the $3 \times 5$ transfer integral matrix from the $d$ orbitals of the $i$th TM ion to the $p$ orbitals of the ligand shared by the $i$th and $j$th octahedra defined in the coordinate for the $i$th octahedron; $\hat{\tau}^{pd}_{i;ij}$ is given by the Slater-Koster parameters $V_{pd\sigma}$ and $V_{pd\pi}$~\cite{harrison}.  
$\hat{\cal R}_{ij}(\phi)$ is the $3 \times 3$ matrix defined by $\hat{\cal R}_{ij}(\phi) = \hat{R}_{i}^\top(\phi) \hat{R}_{j}(\phi)$, where $\hat{R}_i(\phi)$ represents the rotation of the $i$th octahedron.

An important feature of the resultant $d$-$d$ transfer integrals is their spatial anisotropy depending on the bond directions, owing to the hybridization between the different $d$ orbitals induced by the {\it BX}$_6$ rotations. 
The $d$ sites are divided into four kinds of sites termed $B_1$-$B_4$ in the unit cell; $B_1$ and $B_2$, $B_3$ and $B_4$ are respectively connected by the glide symmetry [the glide plane is drawn in Fig.~\ref{fig1}(c)]. 
As shown in Fig.~\ref{fig1}(d), for example, the transfer integral between the $zx$ orbital in $B_1$ [$zx(B_1)$] and the $xy$ orbital in $B_2$ [$xy(B_2)$] in the $(10\bar{1})$ direction becomes nonzero, which is zero in the cubic structure without the GdFeO$_3$-type distortion since the orbitals are orthogonal to each other, and is larger than that between the $yz(B_1)$ and $xy(B_2)$ orbitals in the $(101)$ direction.
On the other hand, on the $B_2$-$B_1$-$B_2$ bonds, the magnitudes of the electron transfers in the $(101)$ and $(10\bar{1})$ directions are switched with each other as shown in Fig.~\ref{fig1}(e), reflecting the glide symmetry which connects the $B_1$ and $B_2$ sites. 
These yield the anisotropic transfer integrals depending on the bond directions within the $ac$ plane.

The onsite Coulomb interactions between the $d$ electrons are introduced in the conventional manner as 
\begin{eqnarray}
{\cal H}_{\rm int} 
&=& U \sum_{i \beta} n_{i \beta \uparrow} n_{i \beta \downarrow} + \frac{U'}{2} \sum_{i \beta \neq \alpha'} n_{i \beta} n_{i \beta'} \notag \\
&+& J \sum_{i \beta > \beta' \sigma \sigma'} c^\dagger_{i \beta \sigma} c^\dagger_{i \beta' \sigma'} c_{i \beta \sigma'} c_{i \beta' \sigma} \notag \\
&+& I \sum_{i \beta \neq \beta'} c^\dagger_{i \beta \uparrow} c^\dagger_{i \beta \downarrow} c_{i \beta' \downarrow} c_{i \beta' \uparrow},
\end{eqnarray}
where $U$ and $U'$ represent the intra- and inter-orbital Coulomb interactions, respectively, $J$ is the Hund coupling, and $I$ is the pair hopping interaction.
Here we do not take into account the spin-orbit coupling which is not essential for the present mechanism.

We analyze the effective five-orbital Hubbard model defined by ${\cal H} = {\cal H}_0 + {\cal H}_{\rm int}$ within the Hartree-Fock approximation, where the mean fields $\langle c^{\dagger}_{i\beta \sigma} c_{i\beta' \sigma} \rangle$ are self-consistently determined. 
Using the Hartree-Fock eigenenergies and eigenstates, the conductivity of the spin current with the spin parallel to the AFM moment along the $\mu$ axis with respect to an electric field parallel to the $\nu$ axis by means of the Boltzmann transport theory with the relaxation-time approximation is given by 
\begin{equation}
\chi_{\mu\nu} = \frac{-e \tau}{2 N V} \sum_{\bm k m \sigma} \sigma v_{{\bm k} m \sigma}^{\mu} v_{{\bm k} m \sigma}^{\nu} \left[ -\frac{\partial f(\epsilon_{{\bm k}m\sigma})}{\partial \epsilon_{{\bm k}m\sigma}} \right],
\end{equation}
where $f(\epsilon_{{\bm k}m\sigma})$ is the Fermi distribution function for the Bloch eigenstate with the wave vector $\bm k$, the band index $m$, and the spin $\sigma(=\pm1)$.
$v_{{\bm k} m \sigma}^{\mu}$ is the $\mu$ component of the group velocity of the wave packet centered on the state $\left| \bm k m \sigma\right\rangle$, $\tau$ is the transport relaxation time, $V$ is the volume of the unit cell, and $N$ is the total number of the unit cells.
Note that the spin current is conserved in the present model without the spin-orbit coupling.
Here our treatment is equivalent with the linear response theory by the Kubo formula when the relaxation time is long enough~\cite{gonzalez}. 

\begin{figure}[t]
\begin{center}
\includegraphics[width=1.0\columnwidth, clip]{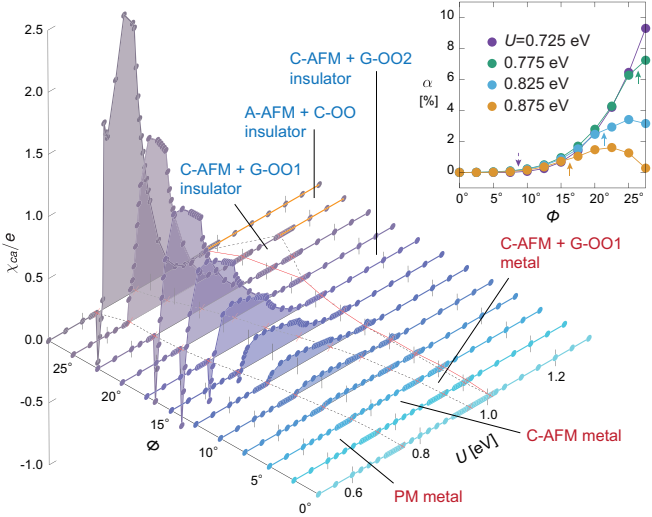}
\end{center}
\caption{
Spin current conductivity $\chi_{ca}$ on the $U$-$\phi$ plane. 
The black broken lines on the basal plane indicate the phase boundaries between the different phases. 
The red solid line stands for the MI transition.
The thick orange line represents the region where the A-AFM + C-OO phase is stable.
Inset shows $\phi$ dependences of the charge-spin current conversion rate $\alpha$ in the C-AFM metallic phases.
The broken and solid arrows indicate the phase boundaries between the PM and C-AFM phases and between the C-AFM and C-AFM + G-OO1 phases, respectively.
}
\label{fig2}
\end{figure}
In the following calculation, we choose the model parameters typical of perovskite $3d$-TM oxides as $V_{pd\sigma}=2$ eV, $V_{pd\pi}=1$ eV, $\Delta_{\rm cf}=3$ eV, and $\Delta_{\rm ct}=5$ eV~\cite{mizokawa}. 
As for the Coulomb interaction terms, we adopt the relations $U=U'+2J$ and $J=I$~\cite{kamimura}. 
While the model is known to exhibit a variety of spin and orbital orders depending on the electron filling $n$ (the electron number per site on average) and other parameters~\cite{mizokawa}, we find that the C-AFM order can make an off-diagonal component of the spin current conductivity, $\chi_{ca}$, nonzero; hence, below we show the results for $n=2$, where the C-AFM order was obtained under electron correlations and the GdFeO$_3$-type distortion~\cite{mizokawa}, by changing $U$ while the ratio as $U'=2U/3$ and $J=U/6$, and the rotation angle of the {\it BX}$_6$ octahedra $\phi$.
We note that $\chi_{ca}$ becomes nonzero under the C-AFM order irrespective of $n$.
The $\bm k$-space mesh, equivalent to $N$, is chosen as $16 \times 10^3$ and $10^6$ for evaluating the mean fields and the spin current conductivity, respectively.
We set the relaxation time as $1/\tau = 1$~meV and the lattice constants to unity. 

\begin{figure}[t]
\begin{center}
\includegraphics[width=1.0\columnwidth, clip]{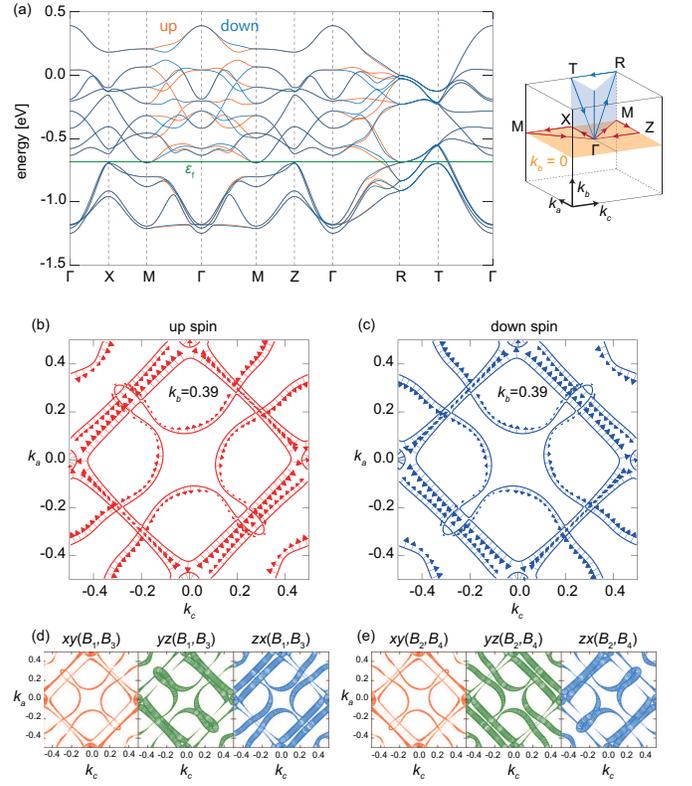}
\end{center}
\caption{
(a) Energy band structure in the C-AFM metallic phase at $(U, \phi)=(0.725 \ {\rm eV}, 25^\circ)$.
The Fermi energy is denoted by $\varepsilon_{\rm f}$.
The right panel shows the symmetric lines in the first Brillouin zone.
(b) and (c) show the group velocities of the up- and down-spin electrons, respectively, on the Fermi surfaces at $k_b=0.39$.
$k_c$ and $k_a$ stand for the coefficients of the reciprocal vectors ${\bm b}_c$ and ${\bm b}_a$, respectively.
The length of the red and blue arrows represents the magnitude of the group velocity on each $\bm k$ point. 
(d) and (e) are the relative weights of the $t_{2g}$ orbitals on the up- and down-spin Fermi surfaces, respectively.
The areas of the red, green, and blue circles represent the magnitudes of the $xy$ (left), $yz$ (middle), and $zx$ (right) components on (b) $(B_1, B_3)$ and (c) $(B_2, B_4)$ sublattices.
}
\label{fig3}
\end{figure}
Figure~\ref{fig2} shows the off-diagonal spin current conductivity $\chi_{ca}$ as a function of $U$ and $\phi$, together with the phase diagram on the basal plane.
In the absence of the GdFeO$_3$-type distortion ($\phi=0$), when $U$ is increased from the paramagnetic (PM) metallic phase, a phase transition to the C-AFM metallic phase occurs at $U \simeq 0.75$ eV. 
By increasing $U$ further, two kinds of G-type orbital ordering (G-OO1 for $0.95 \ {\rm eV} \lesssim U \lesssim 1.01 \ {\rm eV}$ and G-OO2  for $U \gtrsim 1.01$ eV) are stabilized coexisting with the C-AFM order.  
In the two orbital ordered phases, nearly one electron commonly occupies the $xy$ orbital at each site, and the other electrons alternately occupy $\frac{1}{\sqrt{2}} (yz + zx)$ and $\frac{1}{\sqrt{2}} (yz - zx)$ in G-OO1 while they occupy $zx$ and $yz$ in G-OO2~\cite{mizokawa}.
Near the phase transition between the G-OO1 and G-OO2 phases, the system undergoes a MI transition. 
These C-AFM phases are robust against the increase of $\phi$ except for at $\phi \gtrsim 25^\circ$ in the large $U$ region, where the A-type AFM  (A-AFM) phase accompanied by the C-type orbital order (C-OO) of the $yz$ and $zx$ orbitals is stabilized instead.

As shown in Fig.~\ref{fig2}, $\chi_{ca}$ is constantly zero at $\phi=0$, irrespective of the value of $U$, while it turns nonzero in the presence of the GdFeO$_3$-type distortion in the C-AFM metallic phases.
This means that in this region a charge current along the $a$ axis is converted into a spin current along the $c$ axis.
We note that the conductivity tensor is symmetric, $\chi_{ca}=\chi_{ac}$, with vanishing diagonal elements, $\chi_{aa}=\chi_{cc}=0$.
The inset of Fig.~\ref{fig2} shows the $\phi$ dependences of the charge-spin current conversion rate defined by $\alpha \equiv (2e/\hbar) (\chi_{ca}/\sigma_{aa})$ with the electrical conductivity $\sigma_{aa}$ calculated within the same scheme~\cite{zelezny, naka}.
$\alpha$ basically increases with $\phi$ up to around $6$ $\%$ at $\phi=25^\circ$ in the C-AFM metallic phase, which is comparable to that observed in Pt originating from the spin Hall effect due to the strong spin-orbit coupling~\cite{wang}. 

Let us investigate the microscopic mechanism of the off-diagonal spin current response.
First, we show the energy band structure in the C-AFM phase at $(U, \phi) = (0.775 \ {\rm eV}, \ 25^\circ)$, where the Fermi energy resides in the $t_{2g}$ bands.
Figure~\ref{fig3}(a) shows the up- and down-spin electron bands along the symmetric lines illustrated in the inset. 
A spin splitting appears in the general $\bm k$ points except for the planes $k_c=0$, $\pm0.5$ and $k_a=0$, $\pm0.5$ in the Brillouin zone.
This is a consequence of the glide symmetry breaking by the collinear AFM order (note that the spin-orbit coupling is absent in our model), similarly discussed in different systems~\cite{naka, hayami, gonzalez, hayami2}. 

Next, we show the up- and down-spin Fermi surfaces in Figs.~\ref{fig3}(b) and \ref{fig3}(c), respectively, together with the group velocities $\bm{v}_{\bm{k}n\sigma}$ along them. 
We show the results at $k_b=0.39$ since the Fermi surfaces reside near $k_b \simeq 0.5$ [see Fig.~\ref{fig3}(a)].
The up-spin Fermi surfaces are mainly composed of the one-dimensional-like curves along $(101)$ and the two-dimensional rhombi. 
There, the group velocities are relatively large on the parts of the Fermi surfaces parallel to the $(101)$ direction. 
These features imply that, when an electric field is applied along the $(100)$ direction, the up-spin electrons drift to the $(\bar{1}01)$ direction. 
On the other hand, the down-spin Fermi surfaces, which correspond to the mirror images of the up-spin ones with respect to the $k_a$-$k_b$ plane, carry the down-spin electrons along the $(\bar{1}0\bar{1})$ direction under the $(100)$ electric field.
As a result of this spin-dependent anisotropy, the spin current flows along $(010)$ direction perpendicular to the electric field. 

Furthermore, we can pin down the $d$-orbital component most responsible for the spin-dependent transport. 
Figures~\ref{fig3}(d) and \ref{fig3}(e) show the relative weights of the $t_{2g}$ orbitals on the ($B_1$, $B_3$) sublattice in the up-spin Fermi surfaces, and those on the ($B_2$, $B_4$) sublattice in the down-spin Fermi surfaces, respectively; the other components are smaller and therefore not shown.
The $zx$($yz$) orbitals on the $B_1$($B_2$) and $B_3$($B_4$) sites, where the majority spin is up(down), compose the Fermi surfaces parallel to $(101)$[$(\bar{1}01)$], as mentioned above, dominating the anisotropic spin transport.
These are indeed consistent with the real-space anisotropic transfer integrals due to the GdFeO$_3$-type distortion shown in Figs.~\ref{fig1}(d) and \ref{fig1}(e); the up(down)-spin on the $B_1$($B_2$) sites tend to hop in the $(\bar{1}01)$[$(101)$] direction having a larger transfer integral than the $(101)$[$(\bar{1}01)$] direction. 

A similar anisotropic spin transport owing to the combination of a collinear AFM spin structure and site-dependent anisotropic transfer integrals was proposed for quasi-two-dimensional $\kappa$-type organic conductors~\cite{naka}.
In that case molecular dimers have two different orientations leading to dimer-dependent anisotropic transfer integrals. 
From the viewpoint of crystalline symmetry, these two systems belong to the same space group {\it Pnma}, which also support the similarity; the C-AFM order here corresponds to the AFM structure in the $\kappa$-type organic system, both breaking the in-plane glide symmetry. 
We note that a similar mechanism without the spin-orbit coupling was discussed for other materials~\cite{hayami, gonzalez, yuan, hayami2}.
Our work here illuminates spin current generation feasible in ubiquitous materials based on a microscopic model calculation.

Finally, we discuss possible experimental detections of the present spin current generation in perovskites.
One of the candidates is CaCrO$_3$ containing Cr$^{4+}$ ions with nominally $(3d)^2$ electron configuration, which shows the C-AFM metallic state below 90 K~\cite{komarek, bhobe, streltsov}. 
The Cr-O-Cr bond angle is $160^\circ$, which corresponds to the {\it BX}$_6$ rotation angle $\phi \approx 12^\circ$.
Within our model calculation, the spin-charge conversion ratio $\alpha$ is estimated up to around $0.5 \ \%$.
The C-AFM state is also observed in a series of vanadates {\it A}VO$_3$ with {\it A}=La-Y, also having the $(3d)^2$ electron configuration and $\phi \approx 14^\circ$-$22^\circ$, below around 100-150 K~\cite{zubkov, sawada, miyasaka, motome}. 
These values of $\phi$ correspond to $\alpha = 1$-$4 \ \%$ in our calculation.
Although the C-AFM phases in these compounds are generally insulating due to the coexistence of the G-type OO, the carrier doping by substitution of the {\it A} ion reduces the OO and stabilizes the metallic phase~\cite{miyasaka2}.
In addition, we note that the present mechanism works in a wide range of the electron filling once the C-AFM order is stabilized.
For example, manganites {\it R}$_{x}${\it A}$_{1-x}$MnO$_3$ which are known to show the C-AFM state~\cite{tokura, dagotto, kajimoto} are also candidates.

In summary, we have shown that perovskites can serve as a spin current generator by applying an electric field, which originates from a cooperation of the GdFeO$_3$-type lattice distortion ubiquitously seen in a wide range of perovskites and C-type AFM ordering. 
We note that this is the case including not only $3d$- but also $4d$- and $5d$-TM ions. 
As the present mechanism can also be interpreted as another type of ``spin-orbit coupling'' which brings about the spin-split band, its cooperative effect with the conventional spin-orbit coupling~\cite{naka2, smejkal, gonzalez} is an intriguing future issue.
Another issue is spin current generation by applying a thermal gradient to the C-AFM Mott insulating state. 
Here the spatially anisotropic electron transfers lead to the anisotropic exchange interactions, which will give rise to a spin-dependent transport of the magnon, as demonstrated in Ref.~[\onlinecite{naka}]. 

\begin{acknowledgments}
The authors would like to thank Y. Fuseya, T. Katsufuji, T. Mizokawa, and M. Mochizuki for valuable comments and discussions. 
This work is supported by Grant-in-Aid for Scientific Research, No. JP19K03723,  No. JP19K21860, JP19H05825, JST-CREST (No. JPMJCR18T2), and the GIMRT Program of the Institute for Materials Research, Tohoku University, No. 19K0019.
\end{acknowledgments}



\end{document}